\documentclass[letterpaper]{article} % DO NOT CHANGE THIS
\usepackage{aaai2026}  % DO NOT CHANGE THIS
\usepackage{times}  % DO NOT CHANGE THIS
\usepackage{helvet}  % DO NOT CHANGE THIS
\usepackage{courier}  % DO NOT CHANGE THIS
\usepackage[hyphens]{url}  % DO NOT CHANGE THIS
\usepackage{graphicx} % DO NOT CHANGE THIS
\usepackage{natbib}  % DO NOT CHANGE THIS AND DO NOT ADD ANY OPTIONS TO IT
\usepackage{caption} % DO NOT CHANGE THIS AND DO NOT ADD ANY OPTIONS TO IT
\usepackage{amsmath}
\usepackage{cleveref}
\usepackage{amsfonts} 
\author{
    Jinjian Liu\textsuperscript{\rm 1}\equalcontrib,
    Yichuan Wang\textsuperscript{\rm 1}\equalcontrib,
    Xinxi Lyu\textsuperscript{\rm 2},
    Rulin Shao\textsuperscript{\rm 3}, \\
    Joseph E. Gonzalez\textsuperscript{\rm 1},
    Matei Zaharia\textsuperscript{\rm 1},
    Sewon Min\textsuperscript{\rm 1}
}
\affiliations{
    \textsuperscript{\rm 1} University of California, Berkeley \\
    \textsuperscript{\rm 2} University of Illinois Urbana–Champaign \\
    \textsuperscript{\rm 3} University of Washington \\
    leifmax@berkeley.edu, yichuan\_wang@berkeley.edu, seanlyu2@illinois.edu, rulins@cs.washington.edu, \\
    jegonzal@cs.berkeley.edu, matei@berkeley.edu, sewonm@berkeley.edu
}

\usepackage{booktabs}
\usepackage{xspace}
\usepackage{algorithm}
\usepackage{algorithmic}
\usepackage{newfloat}
\usepackage{listings}
\usepackage{xcolor}
\usepackage{todonotes}
\newcommand{\sys}{\textsc{DS Serve\xspace}}%
\newif\ifcomments
\commentstrue
\ifcomments
    \newcommand{\sewon}[1]{{\color{purple!70}[sewon: #1]}}
    \newcommand{\yichuan}[1]{{\color{blue}[yichuan: #1]}}
    \newcommand{\jinjian}[1]{{\color{teal}[jinjian: #1]}}
\else
    \newcommand{\sewon}[1]{}
    \newcommand{\yichuan}[1]{}
    \newcommand{\jinjian}[1]{}
\fi
\newcommand{\myskip}[1]{}

\newcommand{\mysmallerfontsize}{\footnotesize} 

\newfloat{listing}{tb}{lst}{}
\floatname{listing}{Listing}

\title{\sys: A Framework for Efficient and Scalable Neural Retrieval}
\usepackage{bibentry}

\begin{document}

\maketitle

\begin{abstract}
We present \sys, a framework that transforms large-scale text datasets---comprising half a trillion tokens---into a high-performance neural retrieval system. \sys\ offers both a web interface and API endpoints, achieving low latency with modest memory overhead on a single node. The framework also supports inference-time tradeoffs between latency, accuracy, and result diversity. We anticipate that \sys\ will be broadly useful for a range of applications such as large-scale retrieval-augmented generation (RAG), training data attribution, training a search agent, and beyond.  
\end{abstract}

\begin{links}
    \link{Code}{github.com/Berkeley-Large-RAG/RAG-DS-Serve} \\
    %\link{Dataset}{huggingface.co/datasets/alrope/CompactDS-102GB} \\
    \link{Demo URL}{http://api.ds-serve.org:30888/ui}
\end{links}

\section{Introduction}

Neural retrieval over large-scale text datasets comprising nearly a trillion tokens has become central to modern machine learning, powering applications from retrieval-augmented generation (RAG) to training data attribution and curation.
Yet deploying such systems at this scale remains difficult: high latency and memory requirements make them costly and often impractical for fast inference or interactive use. For example, a dataset from CompactDS---a pre-training dataset with half a trillion tokens~\cite{compactds}---produces two billion 768-dimensional vectors, with raw embeddings exceeding 5TB, posing a significant challenge for existing retrieval frameworks. Current systems typically optimize for either accuracy or efficiency, but rarely both, and often require distributed infrastructure, creating barriers for researchers and practitioners with limited resources.

To this end, we present \sys, a framework that transforms massive text datasets into a neural retrieval system designed to run efficiently on a single node. \sys\ leverages approximate nearest neighbor search, and is capable of handling billions of vectors (e.g., 2B in our deployment). It further supports exact and diversity-based reranking, enabling inference-time tradeoffs among accuracy, latency, and diversity.
On CompactDS data, \sys\ delivers subsecond latency with modest memory overhead ($\approx$200GB RAM), demonstrating scalability without distributed infrastructure. \sys\ provides both a web interface and API endpoints, easing integration across diverse workflows.

\begin{figure}[t]
    \centering
    \includegraphics[width=\columnwidth,
    % height=0.12\textheight
    ]{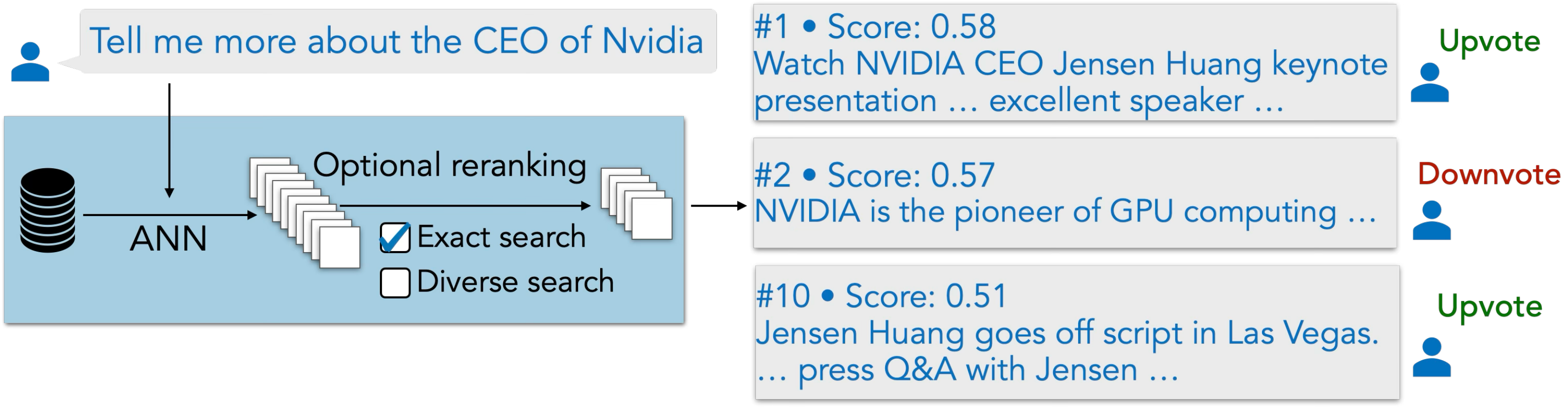}
    \vspace{-1.7em}
    \caption{
    \sys\ converts a large dataset into a neural retrieval system: a query $q$ retrieves relevant text via ANN (DiskANN or IVFPQ), optionally reranks with Exact and/or Diverse Search, and returns the top-$k$ chunks with voting for feedback.
}
    %\vspace{-0.1cm}
    \label{fig:placeholder}
\end{figure}

In summary, \sys\ enables building \emph{fully controllable} in-house retrieval systems over arbitrary \emph{large-scale} datasets on a single node, with accessible interaction via a web interface and API endpoints. 
\sys\ is broadly applicable to large-scale RAG, training data attribution, search agent training, and other retrieval-intensive applications.

%\yichuan{all above(intro and abstract) looks good for me}

%Our contributions are threefold: (1) a scalable single-node framework for billion-scale retrieval; (2) mechanisms for inference-time tuning of accuracy, latency, and diversity, and (3) practical accessibility through a web interface with voting features and API endpoints. Together, these contributions make \sys\ broadly applicable to large-scale RAG, training data attribution, search agent training, and other retrieval-intensive applications.

% \yichuan{Do we need to mention "it's a fully controllable, in-house retrieval, which can index arbitrary in-house data", I think it is an exciting point, and we should put it somewhere in the intro? like we share the system optimization to help build such in-house datastore}
%\vspace{-0.5cm}\sewon{I'm a bit worried about vspace between sections}
\section{Description of \sys}
\sys\ is a framework that transforms a large-scale text corpus $\mathcal{D}$ into a high-performance neural retrieval system. % with both a web interface and API endpoints.
% \yichuan{"a web interface and API endpoints" stuff, duplicate with the previous paragraph? or it is ok? I am not that sure} 
This retrieval system takes a user query~$q$ as input and returns a set of text chunks $\{d_1, \ldots, d_k\} \subset \mathcal{D}$, 
where $k$ is a user-specified hyperparameter. % indicating the number of retrieved text chunks. \sewon{Start with the input and the output, e.g., the input is $q$ and the output is $\{d_1, \cdots, d_k\} \in \mathcal{D}$, where $\mathcal{D}$ is the text corpus that is indexed, and $k$ is a hyperparameter. (And once you refine them, you can use consistent notations when describing inner product search and MMR.)}
The retrieval system is primarily based on the DiskANN backbone, with two optional additional modes:
%Users can optionally enable two independent search modes on top of the FAISS ANN backbone:
\emph{Exact Search}, which improves accuracy, and \emph{Diverse Search}, which enhances result diversity.
%When either mode is activated, the corresponding scores (exactness or diversity) are computed, and candidate passages are re-ranked accordingly before the top results are presented.

% \jinjian{
% \section{Overview}
% Given a user query $q$, \sys first retrieves a set of $K$ high-similarity candidate passages using an approximate-nearest-neighbor (ANN) backbone:
% $\{d_1,\ldots,d_K\} \subseteq \mathcal{D}$, where $D$ is the indexed text corpora. However, only a subset of top $\{d_1,\ldots,d_k\}$ passages are displayed according to the user-specified parameter $k$.
% Users can optionally apply two independent search modes on top of the FAISS ANN backbone:
% \emph{Exact Search} that boosts accuracy,
% and \emph{Diverse Search} which diversifies results.
% If either optional mode is enabled, exact or diversity scores will be calculated respectively. Candidate passages will then be re-reranked based on the new scores, before top passages can be displayed.
% } \\

\subsection{\sys\ Backend}
%\jinjian{CompactDS discussion integrated here}
\paragraph{Datastore.}
%\sys\ can process arbitrary in-house datasets at massive scale. Prior work~\cite{massiveds,compactds} has shown that retrieval over large pre-training corpora substantially improves RAG accuracy across diverse tasks such as factoid question answering and reasoning. However, these efforts did not provide an accessible framework for non-experts to build indexes or interact with them through web interfaces or API endpoints. In this paper, we demonstrate \sys\ by constructing an instance on the CompactDS data~\cite{compactds}, which contains 380 billion words spanning web crawl data, Wikipedia, research papers, and beyond.

\sys\ can process arbitrary in-house datasets at massive scale. While prior work~\cite{massiveds,compactds} has shown that retrieval over large pre-training corpora can improve RAG accuracy, such efforts lacked accessible frameworks that allow non-experts to build and interact with indexes. In this paper, We demonstrate \sys\ on CompactDS~\cite{compactds}, a 380B-word corpus (2B vectors) across high-quality sources at a scale far exceeding prior studies ($<50$M vectors~\cite{jin2024long,jin2025search,hu2025hedrarag}) and typical commercial services ($<500$M per namespace~\cite{Turbopuffer2025}).

\vspace{-.3em}
\paragraph{Approximate Nearest Neighbor (ANN) Search.}
Neural retrieval is formulated as a nearest neighbor search that selects the top-$k$ chunks with the highest similarity scores $\operatorname{sim}(\mathbf{q}, \mathbf{d}_i)$, where $\mathbf{q}, \mathbf{d}_i \in \mathbb{R}^h$ are the vector representations of the query $q$ and a candidate chunk $d_i \in \mathcal{D}$, respectively.\footnote{$\mathbf{q}=\mathrm{enc}(q)$ and $\mathbf{d}_i=\mathrm{enc}(d_i)$, where we use Contriever~\cite{izacard2021unsupervised} as the encoder. The function $\operatorname{sim}(\cdot,\cdot)$ denotes the cosine similarity between two embeddings.}

As the datastore $\mathcal{D}$ scales to billions of vectors, efficient nearest neighbor search becomes a central challenge, since linear scans become infeasible.
To address this, we adopt DiskANN \cite{diskann}, a graph-based ANN method that expands a beam of $W$ neighbors while traversing a navigable graph stored primarily on disk. This search implicitly re-ranks candidates in full precision without re-embedding and exposes tunable search complexity $L$ and beam width $W$ so users can explore the accuracy-latency tradeoff. Empirically, DiskANN achieves higher accuracy than IVFPQ \cite{ivfpq} and over 200 end-to-end QPS even with high search complexity. Thus, DiskANN is adopted as the default ANN backend in our framework; users can optionally switch to IVFPQ at their own discretion. 

\vspace{-.3em}
\paragraph{Exact Search.}
While ANN is fast and efficient, it inevitably sacrifices retrieval accuracy, especially at large $|\mathcal{D}|$ where aggressive quantization is required. To mitigate this limitation, \sys\ provides an optional reranking mode based on exact search. In this mode, ANN first retrieves the top-$K$ candidates ($K>k$), which are then reranked using exact similarity scores $\operatorname{sim}(\mathbf{q},\mathbf{d}_i).$
%:$$s_i \;=\; \operatorname{sim}(\mathbf{q},\mathbf{d}_i) \;=\; \frac{\mathbf{q}^{\top}\mathbf{d}_i}{\|\mathbf{q}\|_2\,\|\mathbf{d}_i\|_2}.$$
We use GritLM~\cite{muennighoff2024generative} to compute exact similarity between passages and queries, after which the true top-$k$ passages are returned. Passage vectors are recomputed on the fly during cold start but cached for subsequent queries, typically reducing the latency to below 0.5s when similar queries are posed. As shown in our evaluation, exact search consistently improves accuracy across all tasks (Table 1).

\vspace{-.3em}
\paragraph{Diverse Search.}
Search results often contain substantial overlap, returning nearly identical passages, limiting information breadth. We introduce a \emph{Diverse Search} option 
% \yichuan{can remove this}
, explicitly discouraging redundancy to improve coverage. Concretely, we apply maximal marginal relevance  (MMR) \cite{mmr}: at step $t$, given the already selected set $\mathcal{S}$, each remaining candidate $i$ receives a score:$$\lambda\, \operatorname{sim}\!\bigl(\mathbf{q},\mathbf{d}_i\bigr) \;-\; (1-\lambda)\,\max_{j\in\mathcal{S}} \operatorname{sim}\!\bigl(\mathbf{d}_i,\mathbf{d}_j\bigr).$$ We find that diverse search substantially improves user experience, though it may not necessarily improve RAG performance. We leave its further evaluation to future work.

%While this approach effectively reduces overlap, it is not universally optimal—accuracy varies across tasks after \emph{Diverse Search}.   Its main value lies in presenting a broader range of results, so users can tune the search to their own discretion.

\subsection{Interface Design}

\sys\ provide a web interface and API endpoints, with inference-time tunable parameters: $k$, two optional post-ANN search modes (Exact and Diverse Search), $n_\text{probe}$ and $\lambda$, controlling the number of chunks retrieved and the trade-offs among accuracy, latency, and diversity.
%\sewon{what each hyperparam means and what tradeoffs they control should be clear from the prev paragraphs, and here, it shouldn't be repeated (at most one-phrase-per- hyperparameter summary.}  
%Every query is logged with its parameters and end-to-end latency for replay and comparison, and  \sewon{This info seems minor to me.}
Users can cast a one-click relevance vote for each chunk, with labels stored for system development and evaluation (Figure~\ref{fig:placeholder}).
%Each passage offers a one-click relevance vote (yes/no) whose labels are stored for later curation (Figure~\ref{fig:placeholder}).

% \yichuan{@Jinjian pleasRefe add both latency and accuracy?}

\begin{table}[t]
\centering
\mysmallerfontsize
\setlength{\tabcolsep}{3pt}
\resizebox{\linewidth}{!}{%
\begin{tabular}{l c cc ccc}
\toprule
& \multicolumn{1}{c}{No \sys}
& \multicolumn{2}{c}{\sys}
& \multicolumn{3}{c}{\sys\ w/ Exact} \\
\cmidrule(r){2-2}\cmidrule(lr){3-4}\cmidrule(lr){5-7}
Task & \multicolumn{1}{c}{Acc} & Acc & $t$ & Acc & $t$ & \(t_{\text{cache}}\) \\
\midrule
MMLU     & 68.9 & 73.5 & 0.17 & 73.7 & 16.44 & 0.30 \\
MMLU Pro & 39.8 & 47.5 & 0.19 & 49.4 & 16.54 & 0.32 \\
AGI Eval & 56.2 & 56.2 & 0.21 & 58.3 & 15.03 & 0.34 \\
MATH     & 46.9 & 50.0 & 0.18 & 53.1 & 16.51 & 0.33 \\
GPQA     & 29.9 & 31.7 & 0.17 & 36.6 & 16.57 & 0.32 \\
\bottomrule
\end{tabular}%
}
\caption{
Evaluation of \sys\ on five established benchmarks.
`{\em Acc}' is accuracy (\%), and \(t\) is end-to-end \emph{retrieval} latency (s). For Exact Search, we report \(t\) without cache and \(t_{\text{cache}}\) with cache. We use \(K=1000\), \(k=10\), and \(n_{\text{probe}}=256\) for all tasks.
}
\label{tab:results}
\end{table}

% \vspace{-0.5cm}
\section{Evaluation and Application}
We evaluate \sys\ for RAG applications and discuss other potential use cases.

\vspace{.2em}
\noindent
\textbf{RAG.} We evaluate a RAG model based on \sys\ on five established benchmarks (Table~\ref{tab:results}). \sys\ substantially improves accuracy over the baseline with negligible latency overhead, with further gains achieved through exact search. We report the upper bound of exact search latency, but it is often much lower in practice when the same or similar queries were issued previously and benefit from caching.

%first evaluate \sys on popular RAG applications, reporting both accuracy and latency results in Table~\ref{tab:results}.

%\jinjian{OLMoTrace integrated here} 
\vspace{.2em}
\noindent
\textbf{Data Attribution and Curation.} \sys\ can readily be used for training data attribution by indexing the entire pre-training corpus. The closest prior system, OLMoTrace~\cite{olmotrace}, relies on $n$-gram matching, whereas \sys\ considers semantic similarity, making it complementary to or more accurate than OLMoTrace. In addition, \sys\ enables improved data curation through semantic deduplication, decontamination, and customized filtering (e.g., identifying subsets of large datasets relevant to specific queries).

%. OLMoTrace employs an n-gram–based method to map model outputs back to pre-training data. In contrast, \sys leverages semantic search, which can capture high-dimensional similarities beyond surface-level overlap. Furthermore, the system-level optimizations in \sys can be directly applied to building a semantic-search–based pre-training data attribution system. As a result, \sys is a well-suited tool for semantic deduplication, dataset decontamination, and even filtering out unsafe content during pre-training. 

\vspace{.2em}
\noindent
\textbf{Training a Search Agent.}
Search agents are in high demand for applications such as deep research; however, training them is challenging, as rollouts require high-QPS search calls~\cite{jin2025search}, and commercial search engines are costly, slow, and rate-limited. \sys\ addresses these issues by providing a fully controllable search backend, allowing developers to set their own latency-accuracy tradeoffs without incurring costs or rate limits. 
% \sewon{TODO: add citations}

%\sys can be directly applied to training deep research–style agents in post-training\cite{jin2025search} \jinjian{please double-check this citation @yichuan}\yichuan{fixed with a relevant citation}, where the rollout step requires high-QPS search calls. In this setting, \sys provides a low-latency and cost-efficient semantic search endpoint. Beyond this, \sys can also support data curation by identifying subsets of queries relevant to a given task.

\vspace{.2em}
\noindent
\textbf{Pushing the Frontier of Search.}
Commercial web search engines (e.g., Google) are powerful, but they have room for improvement: they perform well on short keyword queries but struggle with long or complex inputs. Vector-based retrieval is more effective in such cases and can complement or even outperform traditional search engines (as shown in \citet{compactds}).
Our voting features also enable collection of real-world labeled data, enabling creation of realistic benchmarks and training data for retrieval research.
%Moreover, our voting features offer a unique opportunity to collect queries from real-world user distributions, enabling the creation of realistic benchmarks and training data to advance retrieval research.

%\jinjian{Traditional search engine comparison moved here} Furthermore, traditional web search engines (e.g., Google) excel at retrieving relevant webpages for short keyword queries, but they struggle with long or complex inputs and their APIs are often costly in both latency and money. By contrast, \sys is designed for free and efficient large-scale semantic search with low latency.

%In addition, by collecting user feedback on retrieval preferences, \sys enables fine-tuning of embedding models with human-in-the-loop contrastive learning.

%Finally, for the vector search community, \sys provides a unique opportunity to collect queries from real-world user distributions, allowing the construction of realistic benchmarks to advance ANN research.

\bibliography{aaai2026}

\end{document}